\documentclass[preprint,authoryear,12pt]{elsarticle}

\usepackage{float}
\restylefloat{table}
\usepackage{graphicx}
\usepackage[colorlinks,urlcolor=blue,citecolor=blue,linkcolor=blue]{hyperref}
\usepackage{verbatim}
\usepackage{color, colortbl}
\definecolor{Gray}{gray}{0.9}

\usepackage{amssymb}

\journal{Astronomy and Computing}

\begin{document}

\begin{frontmatter}



\title{Comparison of Strong Gravitational Lens Model Software II. HydraLens: Computer-Assisted Strong Gravitational Lens Model Generation and Translation \tnoteref{t1}}
\tnotetext[t1]{HydraLens is available at \url{http://ascl.net/1402.023}}

\author{Alan T. Lefor}\ead{alefor@astr.tohoku.ac.jp}

\address{Astronomical Institute, Faculty of Science, Tohoku University, Sendai Japan}

\begin{abstract}
The behavior of strong gravitational lens model software in the analysis of lens models is not necessarily consistent among the various software available,  suggesting that the use of several models  may enhance the understanding of the system being studied. Among the  publicly available codes, the model input files are heterogeneous, making the creation of multiple models tedious. An enhanced method of creating model files and a method to  easily create multiple models, may increase the number of comparison studies. HydraLens  simplifies the creation of model files for four strong gravitational lens model software packages, including Lenstool, Gravlens/Lensmodel, glafic and PixeLens, using a custom designed GUI  for each of the four codes that simplifies the entry of the model for each of these codes, obviating the need for user manuals to set the values of the many flags and in each data field. HydraLens is designed in a modular fashion, which  simplifies the addition of other strong gravitational lens codes in the future.  HydraLens can also translate a model generated for any of these four software packages into any of the other three. Models created using HydraLens may require slight modifications, since some information may be lost in the translation process.  However the computer generated model greatly simplifies the process of developing multiple lens models. HydraLens may enhance the number of direct software comparison studies, and also assist in the education of young investigators in gravitational lens modeling. Future development of HydraLens will further enhance its capabilities. 

\end{abstract}


\end{frontmatter}



\section{Introduction}

The present time has been referred to as the "Golden Age" of Precision Cosmology \citep{CoeThesis}. Strong gravitational lensing data is a rich source of information about the structure and dynamics of the universe, and these data are contributing significantly to this notion of precision cosmology.  Strong gravitational lens studies are highly dependent on the software used to create the models and analyze the components such as lens mass, Einstein radius, time delays etc. A comprehensive review of available software has been conducted by \cite{Lefor2012}. While many such software packages exist, most studies to date utilize only a single software package for analysis.  Furthermore, most authors of strong gravitational lensing studies use their own software only. More recently, the status of comparative studies of strong gravitational lens models has been reviewed by \cite{Lefor2013}.   

One of the barriers to conducting comparative studies is the heterogeneity of the lens modeling software that currently exists, which includes data input, calculation algorithms, and data output. This heterogeneity is not surprising since all of the software has been independently developed. There are also some common elements among the software being used. This heterogeneity presents one of the greatest barriers to the use of multiple modeling codes in the study of strong gravitational lenses. The data files used by each model code are quite different, and the formats can be confusing for someone wanting to use an unfamiliar lens modeling code. This is a major barrier to comparative studies. Until the present time, software designed to facilitate model entry is only available for Gravlens  \citep{AlfaroWeb}. Using this program is somewhat hampered by the difficulty in compiling it with multiple dependencies. For all other existing lens model software,  lens models files  are entered as a simple free text file, and the user must be careful to count exactly the number of parameters entered on each line and carefully set the values of dozens of numerical flags. Small errors in entry of the file will make the results unpredictable and unusable. 

Some of the software used in lensing studies remains inaccessible to all investigators except the one who developed the software \citep{Sourcecode2013}. In addition to preventing other investigators from duplicating analyses, the lack of availability of software presents another barrier to comparative studies. The Orphan Lens Database \citep{OrphanLensWeb} contains a database of 24 strong gravitational lens modeling software codes. Of these, 16 have been identified as being used in research studies, of which five (Mirage, ZB, WSLAP, SaWLens and GLEE) are not publicly distributed and are used almost exclusively by their developers. The remaining 11 strong gravitational lens model software packages are available for download by interested investigators (Lenstool, Lensview, Gravlens, Lensmodel, GRALE, PixeLens, SimpLens, glafic, LensPerfect, IGLOO and GLAMROC). 

\subsection{HydraLens Software}
The software described herein is called "HydraLens" in reference to the multi-headed creature of Greek mythology, and it directly addresses the difficulties associated with writing a lens model for four different, publicly available strong gravitational lens model codes.  HydraLens is freely available, and easy to compile with no-cost compilers, as a single, unified program. There are no dependencies on other software or interfaces. HydraLens facilitates the entry of lens model files for the four codes implemented, by using a simple graphical user interface (GUI) instead of entering multiple parameters in simple text files. Models are entered using a GUI, which has common elements and layout for all four model codes implemented, largely obviating the need for manuals and references. In addition, HydraLens can translate lens model  files among the four  software packages implemented. HydraLens serves two purposes. First, the ability of HydraLens to translate among modeling codes may assist in the conduct of comparative studies. Second, HydraLens is useful for those learning about strong gravitational lens models, enabling straightforward creation of multiple input files. 

\subsection{Organization of this paper}
This paper is organized as follows. In section \S \ref{Methods} we discuss the detailed organization of the HydraLens software. In section \S \ref{models} we discuss the command structure and input files used by each of the four lens model software codes implemented in order to delineate the issues in lens model translation. In section \S \ref{Results}, we discuss the details of lens model generation and translation as implemented in HydraLens. In section \S \ref{Disc} we discuss issues in comparative lens model studies as well as limitations and future development of HydraLens.

\section{Methods} \label{Methods}

\subsection{Strong Gravitational Lens Models}
Each lens model software package uses a different input data format  to construct the model. They do have some features in common, and some are more similar than others. All of them use simple text files as input, but the format of the text files, available functionality and command structures are very dependent on the particular software. Some of the lens model software uses multiple accessory files to provide other data. Each of them has a unique list of commands, with great variability. For example, Lenstool uses a number of commands in the French language. The fact that they use a wide range of flags, with a wide range of meanings, makes writing a lens model file  difficult, especially for the uninitiated. HydraLens was written to simplify the process of creating lens model input files to facilitate direct comparison studies, and to assist those starting in the field. 

\subsection{HydraLens Software Development}
The use of a simple GUI was considered essential in the development of HydraLens, which was implemented in Visual Basic (VB, Microsoft Corp, Redmond WA USA) since VB offers a commonly recognized and easy to code GUI, as well as the fact that VB software runs in nearly any Windows (Microsoft Corp) environment. VB compilers are available at no cost. HydraLens is easily read and modified making HydraLens more generally useful to the astrophysics community. There are extensive comments embedded in the code to allow customization as desired and a user manual supplied.

\subsection{Overview of HydraLens}
For each lens model software implemented, HydraLens has four basic functions: model generation, model write, model parsing and model translation. Each of these  functions is implemented using a modular approach, for each of the four strong gravitational lens software packages in the system. Each of these four basic modules interacts with a common set of data structures that are configured specifically for the lens model software, as shown in Figure 1. 

The model generation function accepts input from the user from a GUI window, and fills in a data structure with the information for that type of model. Alternatively, the data structure can be filled in by a parsing  an existing model by reading each line, then putting the commands and data into the same data structures as the model generation function uses. For example, one might have an existing model for a particular lens system written for Lenstool. This existing model can be read in by HydraLens (model parsing) and then  translated to any or all of the other three model types supported. Once the model information is in the model-specific data structure (through model generation  or through model parsing), it can be written out as a lens model input file, or it can be translated.

\begin{figure}[h]
\centering
\includegraphics[width=9cm, height=6cm]{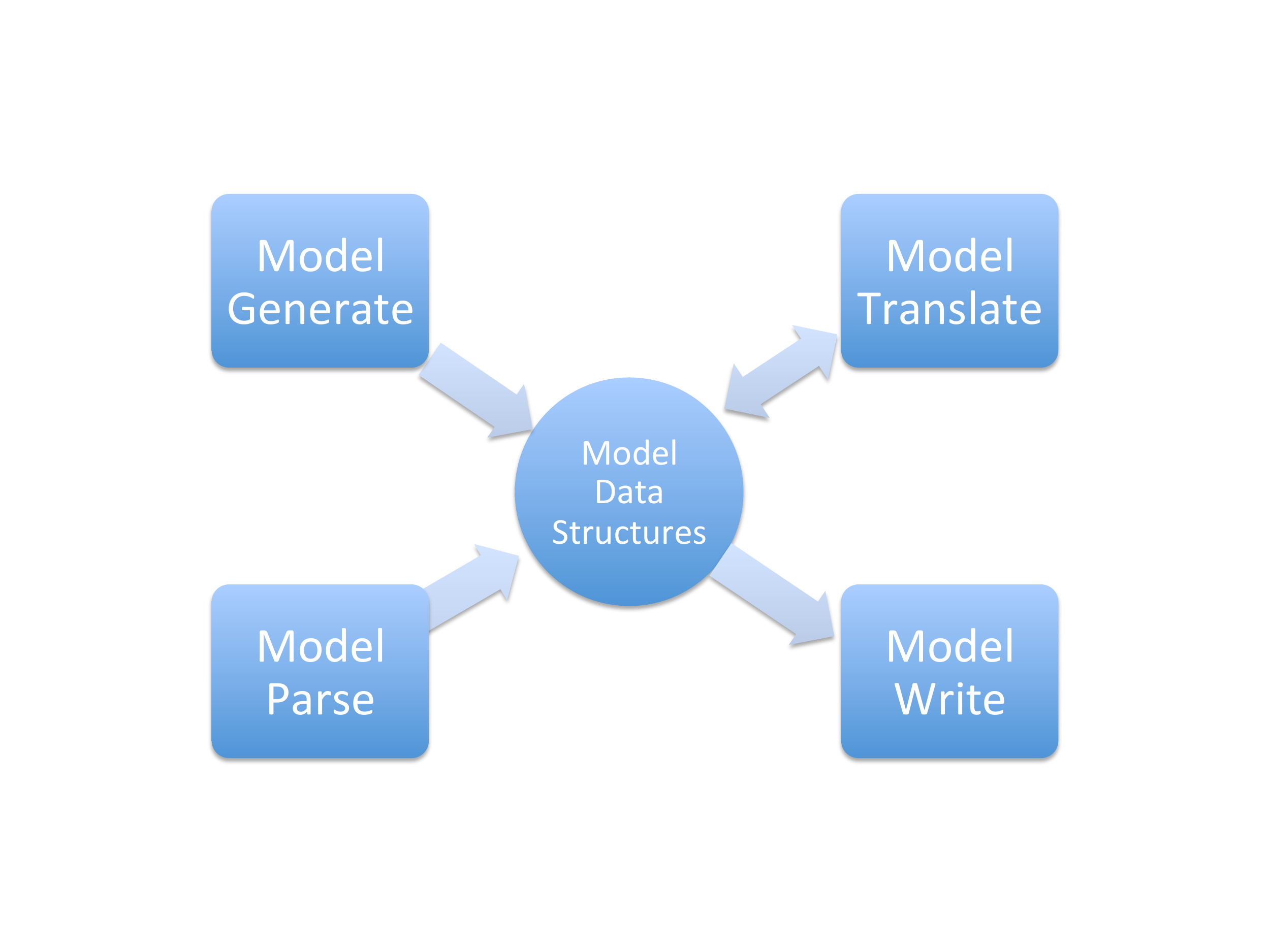}
\caption{Basic data structure of HydraLens showing the interactions of the four modules with the data arrays}
\label{Schema}
\end{figure}

\subsection{Common Parameter Entry}
Most of the information for each model type is entered on a single GUI screen, visible after the user selects the model type to be generated. However, some of the models require the user to enter a number of parameters for each of many lines, such as the .obs file used in glafic which has up to eight parameters for each image for each of the sources entered in glafic. Entering these in a simple text editor is acceptable, but requires the user to be aware of what is typed in each column, with no assistance. For each group of parameters needed, in each program, the software uses a common screen for parameter entry that simply labels each text box, allowing the user to enter text in an appropriately labeled area, then generating the appropriate line for the data file. This parameter entry screen is common to all routines in HydraLens, and greatly simplifies data input (Figure 2). 

\begin{figure}[h]
\centering
\includegraphics[width=9cm, height=6cm]{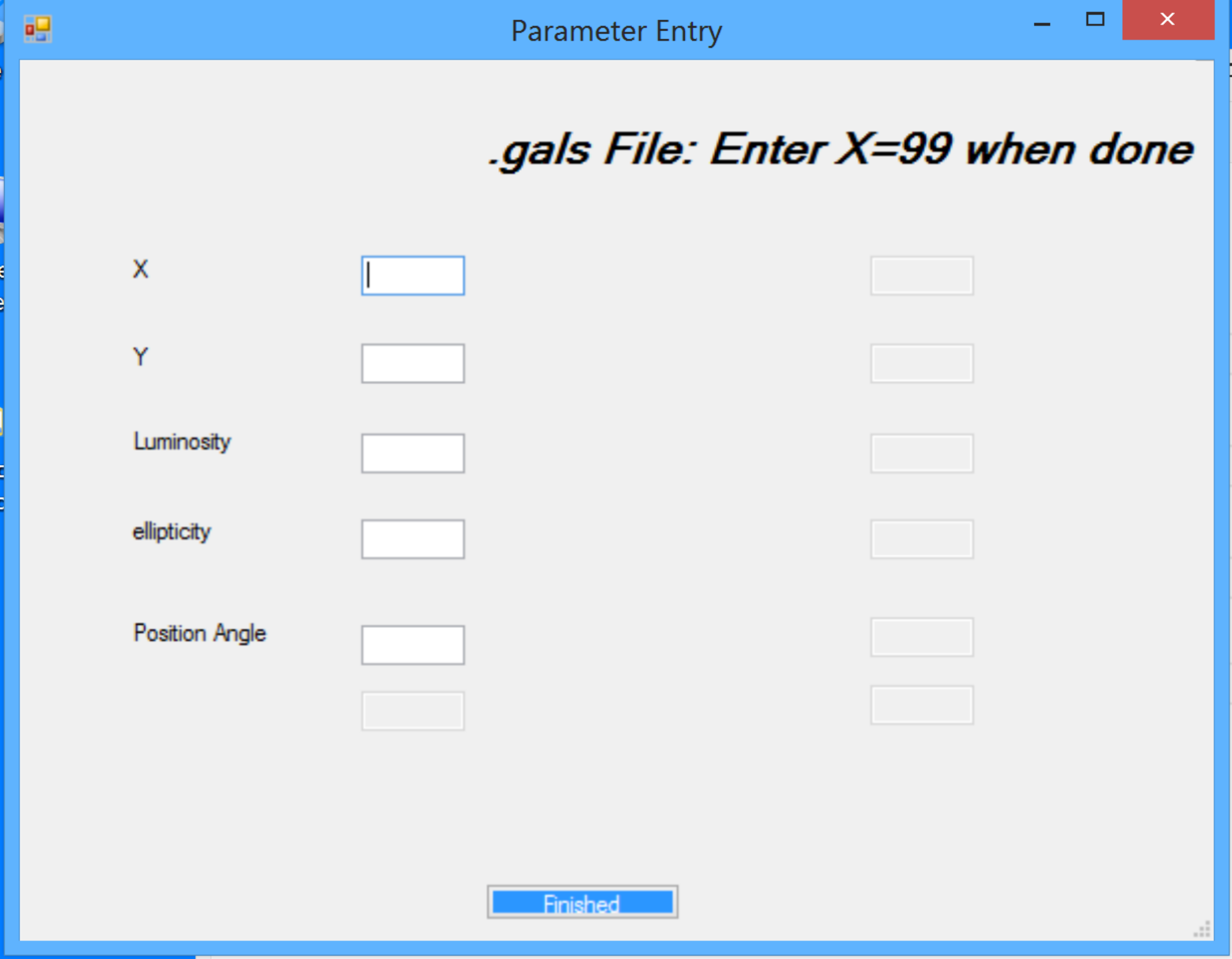}
\caption{Parameter entry is greatly facilitated using a common parameter window, obviating the need to count columns as parameters are entered into labeled text boxes}
\label{Params}
\end{figure}

\subsection{Lens Model Input files}\label{models}
In this section, we discuss the model files for each of the four lens model codes implemented, focusing on  aspects of the input file format important for the generation and translation of the model. Since HydraLens is concerned only with writing lens model files, there is no discussion of output from any of the software. In order to understand the scope of the models available with each of the software packages implemented, it is important to review in some detail the design of each model and the commands available.

\subsubsection{Lenstool}
Lenstool (\url{http://ascl.net/1102.004}) was developed by Kneib, described in 1997 and has undergone several improvements to its algorithms \citep{Jullo2007}. It has been used in many studies in the literature, and uses a combination of light traces mass (LTM, previously known as 'parametric') and non-light traces mass (non-LTM, previously known as 'non-parametric') approaches. Lenstool is available for download as source code and has dependencies on several other software packages to build the software. It is accompanied by a User's Manual \citep{LenstoolWeb}, and there is also a manual written by a third-party which is very useful \citep{LTDummies}. Sample lens model input files are available for download. 

The Lenstool command structure consists of first and second identifiers. The first identifiers are a group of 15 keywords that are basically groups, under which the second identifiers are stated along with values of the parameters. For each of the first identifiers, there is a group of specific second identifiers. Each second identifier is followed by parameters unique to that second identifier such as numerical flags or file names. Each model file does not necessarily contain all of the first identifiers.

The 15 first identifiers in Lenstool include: (Descriptions from the Lenstool Users manual  \citep{LenstoolWeb})

\begin{enumerate}
\item runmode: Reference coordinates can be set (reference), images and arclets (image, arclet) can be defined with the name of input files, a source file (source) can be specified, as well as other second identifiers. 
\item grille: this defines parameters such as number of potential modes, grid mode, polar / rectangular shape of the grid, number of clumps that define the lens potential, and size of the grid.
\item potentiel: Defines the gravitational potential. The profile used is identified by a number, and includes SIS, circular sphere, elliptical sphere, pseudo-elliptical, point mass, PIEMD, plane mass, and NFW profile. For the potential selected, the user specifies a position, ellipticity, angle, and $z_{lens}$. Each different mass model is defined by a numerical flag. Position, mass, ellipticity, velocity dispersion are also set here. 
\item limit: Defines constraints on the potential and is used for optimization.
\item potfile: Default parameters for galaxy scale mass components that account for perturbations to the cluster potential by the galaxies. This includes a filename of the galaxy catalog, mass profile (PIEMD is the default), velocity dispersion, $r_{core}$, $r_{cut}$ among others. 
\item cline: Parameters to compute critical and caustic lines, including the location of the source plane, area to search for critical lines and step between searches. 
\item cosmologie: Specifies the value of constants such as $\Omega_m$, $\Lambda$, $H_0$. 
\item champ: Define size of the field used in some calculations such as dimension of the grid
\item grande: Define representation of the computer deformation of objects
\item observ: Define noise (seeing or Poisson) that is added to a gravitational image.
\item source: Specifies details of the source, including $z_{source}$. 
\item image: Specifies the input data file (object file, with secondary parameter 'multfile') and characteristics of images, multiple images or arclets.
\item cleanlens: Define parameters to retrieve the shape of the source knowing a pixel-frame of the image
\item image: Define characteristics of images, multiple images or arclets
\item fini: Tells Lenstool to stop reading the .par file. This is mandatory.
\end{enumerate}

Lenstool also uses a group of input data files, including:

\begin{description}
\item [Object File] A list of objects characterized by their position, shape and redshift with an integer identifier for each object and six parameters. This format is used for arclets or sources. 
\item [Marker File] A list of marker points in the image plane, with an identifier and xy-coordinate for each. 
\item [IPX Pixel Image File] IPX is a simple format for pixel-images data with a 4 line header. 
\item [FITS pixel image File] This controls the reading of FITS pixel-frames.
\end{description}

A basic Lenstool model  includes the model parameter file (.par file) with primary and secondary identifiers as well as an image file (.cat, in the format of an object file) to define the source images.

\subsubsection{gravlens/lensmodel}
Gravlens/lensmodel  (\url{http://ascl.net/1102.003}) was developed by Keeton, and described in 2001 \citep{GravLens}. These two codes are similar, sharing the same command structure, except that lensmodel adds functionality to the Gravlens kernel. These use a LTM approach to lens models. A paper detailing the mathematics of the mass models in GRAVLENS is also available \citep{Gravlenscat}. The GRAVLENS package is available for download as two executable files, and is accompanied by a User Manual \citep{GravLensManual}. The two executables include gravlens and lensmodel. 

 Sample data files are available for download. Basic commands include:

\begin{description}
\item [Set commands] These are used to set the values of parameters such as $\Omega$, $\Lambda$, $z_{lens}$ and $z_{source}$. There are also a set of flags for gravlens regarding grid format, parity checking, source plane $\chi^2$, tiling and others. In the main data entry screen these values are pre-populated with typical values. 
\item [Data] This command specifies the name of the input data file to use.
\item [Startup] Specifies the number of galaxies for each mass model and the number of mass models, which is followed by a line to specify the mass model selected and the flags for parameters that will be optimized. Once the user selects a particular lens model, the parameters screen opens and the parameters specific to that model are listed with labeled text boxes for entry. Optimization flags are entered separately on the main GUI screen.
\item [Commands] Gravlens has many commands available for use. Some of them require entry of numerous parameters and some are standalone words. The commands allow optimization, varying parameters, data plotting, checking the code, and simple lensing calculations.  Common commands are used to set the type of tiling (grid mode), compute the lensing properties on the specified grid (maketab), check the code (checkder, check mod), create plots of data (plotgrid, plotcrit), and perform simple lensing calculations (calcRein, finding). 
\end{description}

The data file specifies the image data for the lens, including:

\begin{itemize}
\item Number of galaxies
\item Position, $R_{eff}$, PA and e for each galaxy
\item Number of sources
\item Number of images for that source
\item Location, flux, and time delays for each image as well as an identifier
\end{itemize}

A basic gravlens/lensmodel lens model consists of two files. The first is the input file, specifying parameters and  data file name, the mass model to use with optimization parameters, and commands. The second file is the data file which specifies the data for each galaxy and source, as well the images for each of the sources. HydraLens facilitates the creation of both of these files with a GUI interface.

\subsubsection{glafic}
Glafic (\url{http://ascl.net/1010.012}) was developed by Oguri and described in 2010 \citep{Oguri2010}. It has been used in a wide range of studies, and is a LTM approach to strong gravitational lens models, using an adaptive mesh method with increased resolution near the critical curves. Glafic uses functional lens model optimizations with many options. It is available for download as an executable file, and is accompanied by a detailed User Manual \citep{Glaficmanual}. Sample lens models are available for download as well. The structure of glafic is somewhat close in appearance to gravlens.  A glafic input file has three parts. The first part sets the values of various parameters such as $\Omega_m$ and $\Lambda$. The second part defines the lenses, extended sources and point sources. The third part is the list of commands. There is an optional section to define optimizations. 

Parameter settings in glafic:
\begin{description}
\item [Primary parameters] Each of the primary parameters is associated with a flag,  file name, etc. These include $\Omega$, $\Lambda$, $H_0$, $z_{lens}$, output file name, rectangular region of the lens plane, pixel size for extended sources and point sources, and adaptive meshing recursion level.
\item [Secondary parameters] These  include the name of the gals data file, the extended source model arcs file, seed for random number generation, and a number of other parameters and flags that control the behavior of glafic. 
\end{description}

Definition of lenses, extended sources and point sources in glafic:
\begin{description}
\item [lenses] There are 21 different lens mass models in glafic. Each is stored with its name and up to eight parameters. A single lens plane is supported. Most are characterized by a mass scale, x and y coordinates, ellipticity and position angle, and other parameters as needed for the specific mass model. 
\item [extended sources] There are five different extended source types, each of which has up to 8 parameters, including source redshift, coordinates and up to 5 other parameters as indicated. 
\item [point sources] Point sources are stored only as redshift with x and y coordinates. 
\end{description}

Glafic uses a number of secondary files as data for the model, including a galaxy file (galfile.dat), a source file (srcfile.dat), an observation file (obs), and a priors file (prior). Each of these is saved simply as strings based on how many parameters are used in each line.

Data files used by glafic include:

\begin{description}
\item [obs file] File with data of an image of lensed arcs read with command readobs\_point or readobs\_extend (for point sources, extended sources)
\item [gals file] Mass model gals data file (galfile.dat) contains coordinates, luminosity, ellipticity and position angle of each galaxy
\item [src file] Data file used to enter extended source model arcs (srcfile.dat)
\item [prior file] List of priors on parameters, read by 'parprior' command
\item [flux file] Read with the command 'point\_flux', this file contains fluxes for point sources
\item [mask file] Optional file read by 'readobs\_extend' 
\item [sigma file] A list of $\sigma$ values for Markov-Chain Monte Carlo optimization, read by 'mcmc\_sigma'
\end{description}

Optimization Commands in glafic:
\begin{description}
\item [Preparation] read an image of lensed arcs from a file, calculate noise from observed image, read data file for point source optimization, read text file of priors
\item [Setting optimization parameters] Perform model optimization, randomize optimization parameters, calculate a one dimensional $\chi^2$ slice, vary cosmological parameters. 
\end{description}

Commands in glafic:
\begin{description}
\item [Lens properties] Compute various lensing properties for an image, compute Einstein radii for a source redshift, compute mass, write lensing properties to an output file, compute convergences
\item [Extended sources] Write images of lensed extended sources to an output file, calculate total flux, peak count and peak location, and write time delay surfaces
\item [Point sources] Find lensed images for point sources, move source position, compute critical curves and caustics, write mesh pattern, and write time delay surfaces
\item [Other Commands] Other commands are available for composite sources, morsel optimization, and other optimization commands. 
\end{description}

Utilities available in glafic:
\begin{description}
\item [Markov-Chain Monte-Carlo] Read a list of $\sigma$ values for model parameters, perform MCMC calculations, read a resulting chain file. Needs a file of $\sigma$ values.
\item [General Utility functions] Change a parameter value, change optimization flags, moving lens positions, print model parameters or optimization flags, and compute a physical critical surface mass density
\end{description}

A basic glafic lens model includes a parameter / command file (.input) and an image (obs) file.

\subsubsection{PixeLens}
PixeLens (\url{http://ascl.net/1102.007}) was developed by Saha and described in 2006 \citep{Saha2006}. PixeLens is a non-LTM lens model code, and is written in Java which is downloaded as a .jar file and run locally \citep{PixelensWebsite}. It is accompanied by explanatory documentation as well as a tutorial explaining the details of the input file \citep{PixelensWeb}. Sample model files are available on the website. The model files for Pixelens are the simplest among the four codes implemented in HydraLens. Model input can be done through a GUI or through the command line as a batch file that is called when Pixelens in invoked through Java.  The model consists of a group of constants and image data.

\begin{description}
\item [Constants] Pixelens requires an object name, the radius of the mass map in pixels and $z_{lens}$ and $z_{source}$ to be specified. Optionally, one can specify the map symmetry, radius of the mass amp, shear, number of models, Hubble time, minimum steepness, maximum steepness, annular density and cosmological parameters such as $\Omega_m$ and $\Omega_\Lambda$. 

\item [Image Data] Images are given in double or quad format. For each image, one specifies the x and y coordinates as well as the time delay. The redshifts are specified in the first section above. Images must be listed in arrival time order. There is also a 'multi' format used for cluster lenses, useful if there are several source redshifts, or if the image is not a double or quad image. 
\end{description}

A PixeLens model can be entered directly into the Java GUI, or saved as a single text file which contains all the information. HydraLens generates the text file for input to the Java applet.


\section{Implementation}\label{Results}
\begin{figure}[h]
\centering
\includegraphics[width=9cm, height=6cm]{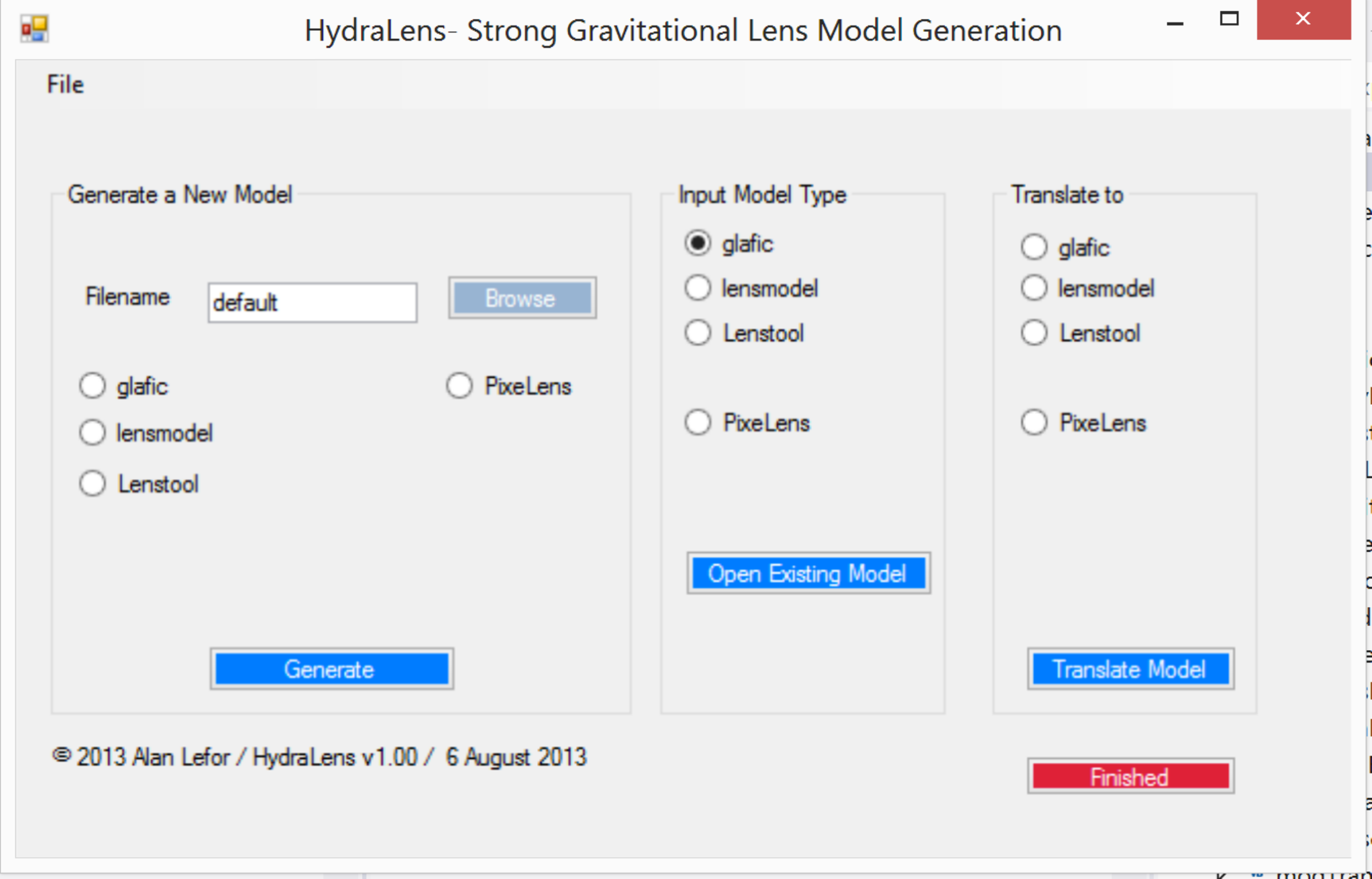}
\caption{The opening screen allows the user to choose to generate a new model, read in an existing model, and translate a model}
\label{OpenScreen}
\end{figure}

When HydraLens is started, the user is presented with an input screen (see Figure \ref{OpenScreen}) to define the name of the model, the directory to store the model and then select the target software from available choices. After selecting the type of model to generate, the user is brought to screens specific for each model.

\subsection{Lens Model Generation}

\subsubsection{Lenstool}
The user first generates any accessory files needed including image file, source file or arclet file. Selecting a button for each file type brings the user to a special screen to build that file type. Upon return to the main model entry screen, the user is presented with detailed entry panels for each of seven commonly used primary parameters in Lenstool, including runmode, grille, potential, limit, cosmologie, image and source. The entry panel for each primary identifier is pre-populated with commonly used values, and parameters are selected using check boxes. When the 'finished' button is pressed, the final lens model is created in the directory selected by the user in the initial screen. 

\subsubsection{gravlens/lensmodel}
HydraLens creates the model file (.input), starting with setting the basic parameter which are on the main gravlens screen pre-populated with typical values. After finalizing the primary parameters, that part of the window becomes invisible, leading the user to enter secondary parameters. The user then specifies secondary parameters as desired. Last, the desired commands are entered from a scrolled list of available commands.  The resulting model file has four sections, including primary parameters, secondary parameters, models / optimizations and commands.  The 'data' command loads the data from a specified file. Once the data command is entered, a button appears on the screen to allow entry of the data file containing the information for each lens galaxy, source, and images for that source. The data file is written, including appropriate comment lines.

\subsubsection{glafic}
After selecting a glafic model, the user selects the type of file to generate (Main model, gals, obs, priors or source) and then goes to a screen specific for that file type. The main model file has a panel for the primary and secondary parameters. Lens models with extended sources and points sources are constructed next followed by entry of desired commands. All available commands are divided into basic calculations, sources, optimization and utilities and selected from lists on the screen.

\subsubsection{PixeLens}
\begin{figure}[h]
\centering
\includegraphics[width=9cm, height=6cm]{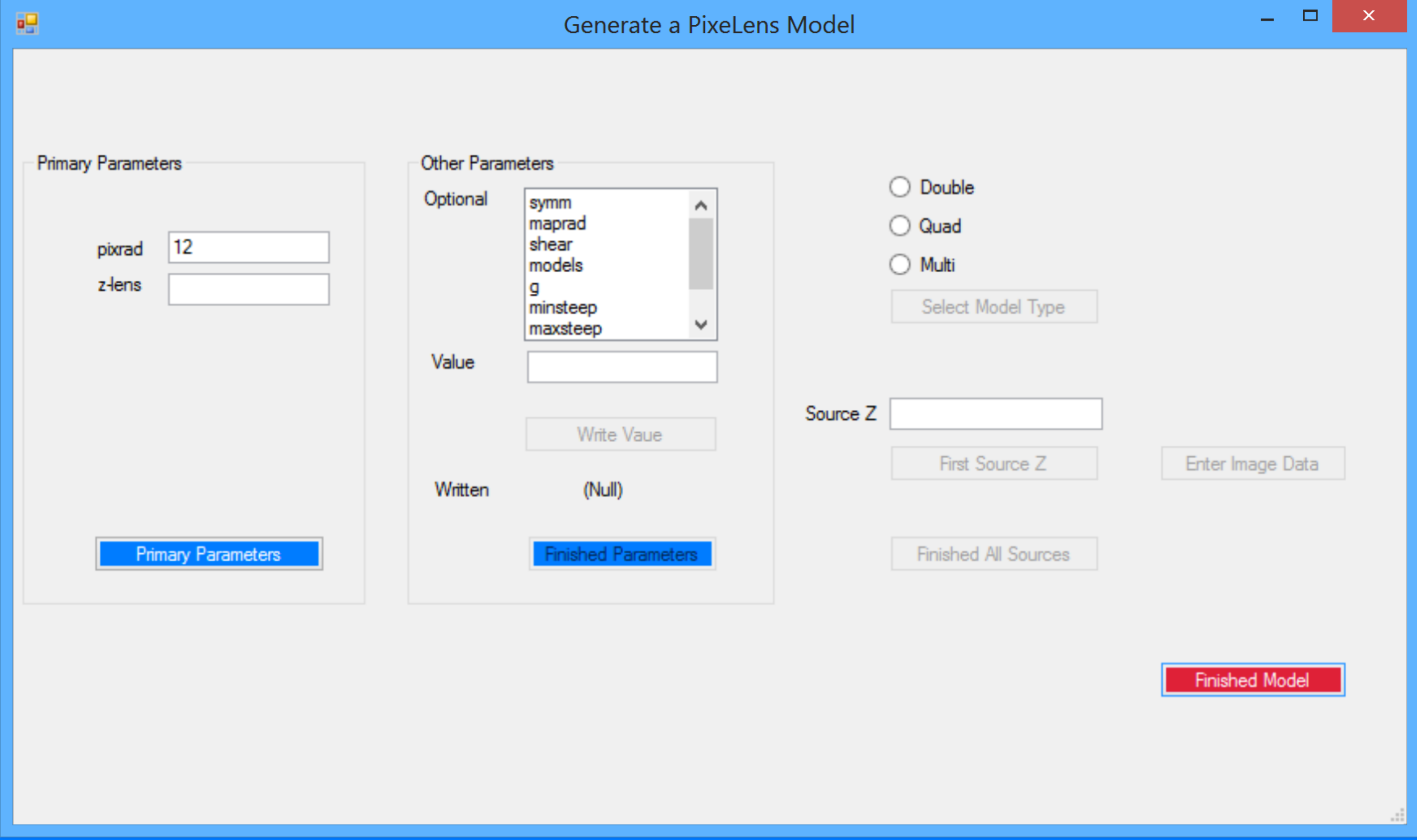}
\caption{The PixeLens model generation entry screen}
\label{PixeLens}
\end{figure}

After selecting a PixeLens model, the user is brought to the PixeLens screen (see Figure  \ref{PixeLens}). The values of required and optional parameters are entered on the left and mage data is entered on the right. Note that the 'action' buttons in the middle and right panels are 'grayed out'. These buttons become active as the user finishes each portion of building the model, to lead them through each step of the process. When the 'Finished' (red, lower right) is pressed, the model is written to the file.

\subsubsection{Completion}
After going through the software specific screens to generate the model, the user is informed that the file has been written and is then brought back to the main screen. At this point the user's only choice is to stop, having generated the model, or to translate it into one of the other three model types. 

\subsection{Lens Model Translation}
The process of translation is performed with no user interaction. After generating a model or specifying the file path to an existing model, and returning to the main HydraLens window, the user selects the software target for translation. The software generates a new model, with an appropriate file name extension and returns to the main HydraLens window so the user can exit. There are four model types supported in HydraLens, so there are 12 different translation modules. Each translation module reads the model that the user just generated from the data structure for that model, translates the parameters and puts them in the data structure for the target model type, then calls the model write routine to write new target model file from the data structure. 

As an example of using HydraLens for translation, a simple model can be easily written and tested in PixeLens, as a way of "rapid prototyping". This simple model can then be translated to models for Lenstool, Lensmodel and glafic in a matter of minutes. The models generated will be functional, but may need modification since many features in glafic, for example, do not exist in PixeLens such as optimizations. The user must then edit the glafic model to set the optimization parameters as desired. In most cases, this is still much faster and simpler than starting with an empty screen in a basic text editing program. Similarly, translation to PixeLens will often result in a simpler model than the original. Another example of information that cannot be translated relates to specific limitations of the codes. In Lenstool, each potential can have its own lens plane, while in the other three codes, only one lens plane is permitted. Thus, translating from Lenstool with such a model necessarily will not include the multiple lens planes. 

It is not possible to transfer all data and/or commands from one type of model to another because of differences in the requirements of each model code. Despite the possible loss of information, the models produced by HydraLens will generally work, and then may need minor modifications to allow for differences in the lens model codes. 

Another difficulty associated with translation is the differences in commands used by the various codes. For example, glafic will calculate the Einstein radius and mass inside the Einstein radius for a Single Isothermal Ellipsoid model by  ignoring the ellipticity. Lensmodel generates an error message when one tries to calculate the Einstein radius for a Single Isothermal Ellipsoid model. Due to the wide range of commands, HydraLens does not translate commands, but rather gives each model a standard group of functioning commands that can be modified by the user. 

The model translation feature offers two important advantages over writing a model using a text editor. First, when creating a new type of model, the image coordinates are easily transferred into the target model, without concern for typing long lists of numbers and counting columns of parameters with proper formatting of the image files. Second, an input file is created with many of the important fields already populated. A minor review of the resulting input file may be necessary, but based on testing to date, the models created will be functional in the target software.  



\section{Discussion} \label{Disc}
HydraLens facilitates creation of strong gravitational lens models for more than one lens model code, in order to facilitate direct comparison studies of strong gravitational lens models. In view of the paucity of direct comparative studies in the literature \citep{Lefor2012, Lefor2013}, HydraLens may help increase the number of future comparative studies by simplifying the process of model development. Additionally, HydraLens may serve an important role in education where students are just starting to use strong gravitational lens modeling codes . HydraLens allows students to easily explore a number of  available software packages. The study of strong gravitational lensing is no longer limited to investigators, but has now extended to being a part of the curriculum in some undergraduate and graduate programs \citep{Seitzlab,Berkeleylab}, as well as being taught to students in specialized intensive education programs \citep{Canary}. The use of lens model software by students may be enhanced by using a tool such as HydraLens to facilitate the writing of lens model input files. 

The use of HydraLens, by both investigators to facilitate comparative studies and by students to use the available software in their studies, is enhanced by the two main functions of HydraLens including lens model file generation and lens model file translation.

\subsection{Limitations}
The major limitation of HydraLens is that it is subject to the unbreakable rule of computing, "garbage-in, garbage-out".  HydraLens cannot write a model in the absence of appropriate input data, and for this reason is referred to as computer-assisted model generation rather than "automatic" model generation. A person totally naive to lens models will not necessarily benefit from HydraLens, without some guidance. Similarly, a person who is an expert at writing lens models for a particular software may not benefit from HydraLens. The people most likely to find HydraLens of value are  those who have begun to write lens models and have some minimal level of experience, or people who are capable with one model software and want to begin using another to conduct direct comparison studies. 

The software described herein is functional and available, and facilitates the writing of lens model files for a variety of available strong gravitational lens model software. For the purpose of writing lens models, HydraLens could be viewed as a specialized text editor. In this role, its major advantage is that the user will rarely need to refer to a reference source for the meaning of most parameters as they are clearly described in the GUI at the time of entry. The input files for lens model software uses simple text files. When writing a model using only a text editor, the user must be very careful about values of flags and parameters, which requires constant reference to users' manuals. HydraLens greatly simplifies that task by entering all fields using a GUI, but the models generated may require some editing. There is no substitute for scientific insight when writing a gravitational lens model. 

In its role as lens model translation software, HydraLens may not always construct a perfect model. Another limitation of model translation is that features vary greatly from one lens model software to another, so that translation may necessitate the loss of some information or capabilities. For example, glafic accepts data on image flux which is not included in Lenstool models. The model created by HydraLens serves as a starting point and eliminates the need for starting the process with a blank piece of paper. Translated models from HydraLens greatly simplifies the tedium of writing an initial model file, especially in regard to image geometry. Generated models are  easily edited since they are all simple text files.


\section{Future Development and Conclusions} \label{Concl}
\subsection{Further Development}
HydraLens is  undergoing further development, especially to improve internal consistency checking within the model. Due to its modular nature, other strong gravitational lens model codes are being built into the system to expand its repertoire of models to generate and translate. These features are being added, and will be included in future releases. 

\subsection{Conclusions}
Previous reviews have shown that there are few comparative studies of strong gravitational lens models in the existing literature \citep{Lefor2012}, yet such comparisons are very important to advance the field. Furthermore, given the differences in results from various strong gravitational lens model codes, such comparitive studies are of great importance \citep{Lefor2013}. 

Barriers to comparative studies include the lack of availability of some software, and the heterogeneity of the input files used in model codes which are available. HydraLens allows the user to enter a lens model with an easy-to-follow GUI rather than entering a tedious text file, for four commonly used strong gravitational lens modeling codes, all of which are freely available for download. Furthermore, HydraLens is capable of translating the data files among the four model codes implemented to allow rapid development and testing of other models for comparison. These features may serve to facilitate direct comparison studies, and also to enhance the educational application of strong gravitational lens model software. Further development is already underway to provide more features and improve the usability of HydraLens.

\section*{Acknowledgements}
The suggestions of the anonymous reviewers are gratefully acknowledged, which facilitated significant  strengthening and clarification of  this manuscript.

\bibliographystyle{model2-names}

\end{document}